\documentclass[prd,preprint,floatfix,showpacs]{revtex4-1}
\usepackage{feynmp,gensymb}
\usepackage{slashed}
\usepackage{calligra}
\usepackage[T1]{fontenc}
\usepackage[pdftex]{color}
\usepackage{latexsym,amssymb,amsmath,amsthm,gensymb}
\usepackage{graphicx,dcolumn,bm,rotating}
\usepackage{bigints,mathtools}
\usepackage[sort&compress]{natbib}
\usepackage{rotating}

\newcommand{\tcr}[1]{\textcolor{red}{#1}}
\newcommand{\tcb}[1]{\textcolor{blue}{#1}}
\def\thalf{ {\textstyle{ \frac{1}{2} }} }
\def\tr{{\rm Tr}}

\definecolor{navyblue}{rgb}{.05,0,.55}
\definecolor{darkgreen}{rgb}{.00,.5,.0}
\newcommand{\tcdg}[1]{\textcolor{darkgreen}{#1}}
\begin{document}

\title {An extension of the standard model
in which\\parity is conserved at high energies}

\author{Kevin Cahill}
\email{kevinecahill@gmail.com}
\affiliation{Department of Physics \& Astronomy,
University of New Mexico, Albuquerque, New Mexico 
87131, USA}
\affiliation{School of Computational Sciences,\\
Korea Institute for Advanced Study, Seoul 130-722, Korea}

\date{\today}

\begin {abstract}
To be compatible with general relativity, every fundamental theory should be invariant under general coordinate transformations including spatial reflection.  This paper describes an extension of the standard model in which the action is invariant under spatial reflection, and the vacuum spontaneously breaks parity by giving a mean value to a pseudoscalar field.  This field and the scalar Higgs field make the gauge bosons, the known fermions, and a set of mirror fermions suitably massive while avoiding flavor-changing neutral currents.  In the model, there is no strong-CP problem, there are no anomalies, fermion number (quark-plus-lepton number) is conserved, and heavy mirror fermions form heavy neutral mirror atoms which are dark-matter candidates.    In models with extended gauge groups, nucleons slowly decay into pions, leptons,  and neutrinos. 
\end {abstract}

\maketitle

\section {Introduction
\label {Introduction} }

Invariance under general coordinate
transformations is the defining principle
of general relativity.  
Although other theories of gravity do 
exist~\cite{Plebanski:1977zz, *PhysRevLett.57.2244, *0264-9381-8-1-009, *0264-9381-8-1-010, *Jacobson:1987yw, *Jacobson:1988yy, *PhysRevD.80.124017, *PhysRevD.89.065017, *PhysRevD.92.043502, *0264-9381-33-2-025011, *Hinterbichler:2011tt},
I will assume in this paper that particle physics
is compatible with general relativity 
and has an action that is invariant
under general coordinate transformations
and in particular under reflection of the spatial coordinates.
The paper describes an extension of the standard model 
in which the action is invariant
under spatial reflection,
and the vacuum spontaneously breaks parity 
by giving a nonzero mean value
to a pseudoscalar field.  
This field and the Higgs field
make the gauge bosons, 
the known fermions, and a set of mirror fermions 
suitably massive
while avoiding flavor-changing neutral currents.
In the model,
gauge fields act on
four-component Dirac fields,
there is no strong-\(CP\) problem,
there are no anomalies,
fermion number (quark-plus-lepton number) 
is conserved,
and heavy mirror fermions
form heavy neutral mirror atoms which
are dark-matter candidates.
In simple grandly unified extensions of 
the model, nucleons slowly decay 
in processes
such as \( p \to \pi^+ + 3 \nu \),
\( p \to e^+ + 4 \nu \), and
\( n \to 3 \nu \)
that conserve \( Q + L \)
but break \( B - L \)\@.
\par
In the standard model, 
the strong and electromagnetic gauge bosons
act on four-component Dirac spinors,
but those of \( SU_w(2) \) act on
two-component left-handed spinors.
This awkward feature breaks parity
and general-coordinate invariance
at the level of the action.
Actions that break parity invite anomalies,
admit the pseudoscalar term
\( \epsilon^{\mu \nu \sigma \tau} F^a_{\mu \nu} F^a_{\sigma \tau} \),
and have a strong-\( CP \) problem.
And in space-times of even spatial dimensions, 
rotational invariance implies invariance
under reflection of the spatial coordinates,
and in space-times of more than three spatial dimensions, 
rotational invariance implies invariance
under reflection of the three spatial coordinates
we know about. 
Most string theories are not
chiral~\cite{Dijkstra:2004cc,*Lebedev:2006kn}\@.
\par
The left-handedness of the weak interactions
led Pati and Salam \cite{PhysRevLett.31.661,
*PhysRevD.10.275}, 
Georgi and Glashow \cite{PhysRevLett.32.438},
and their many followers 
to use two-component left-handed fermion fields
exclusively in their theories of grand unification.
They represented right-handed fermions
as left-handed antifermions, put
quarks and antiquarks in the same multiplet,
and shifted the focus of particle physics 
higher in energy by 12 orders of magnitude.
\par
A theory whose action conserves parity
with gauge fields acting
on four-component spinors 
becomes invariant
under general coordinate transformations
when suitably decorated with tetrads and
Christoffel symbols; it
is free of anomalies and
avoids the problem of
strong-\( CP \) violation.
In the model of this paper
and in its natural grand unifications,  
fermion number \( F \) or
the number of quarks
plus the number of leptons
\begin{equation}
F = Q + L
\label {Fermi number}
\end{equation}
is conserved.  
The model has primary and secondary fermions.
The light, known fermions and the heavy, mirror fermions
are linear combinations of the primary and secondary fermions.
Heavy mirror fermions
form heavy, neutral mirror atoms which 
are dark-matter candidates.
The \(CP\)-breaking phases 
of the heavy fermion sector may 
be enough to explain the excess
of matter over antimatter.
These are some of the advantages of 
theories in which parity is spontaneously broken
and gauge fields act 
on four-component Dirac (or Majorana) spinors.
\par
At energies exceeding those of the
heavy mirror fermions, 
parity is restored in these models.
The lower limits on heavy quarks
are 735 GeV for \( t_3 = \frac{1}{2} \)
and 755 GeV for 
\( t_3 = - \frac{1}{2} \)~\cite{Aad:2014efa}\@.
The lower limits on heavy charged and neutral leptons
respectively are 101.9 GeV and  
80.5--101.5 GeV~\cite{Achard:2001qw}\@.
The restoration of parity may occur
at tens of TeV\@.
\par
The model avoids flavor-changing neutral currents
because the \(3 \times3 \) Yukawa matrices
that couple the scalar and pseudoscalar fields
to the fermions have  
singular-value decompositions
that differ only in their singular values.
For instance, the matrices that give
masses to the three generations
of \( t_3 =1/2 \) quarks
and mirror quarks 
\( U^u \Sigma^u_h V^{u \dagger} \)
and
\( U^u \Sigma^u_p V^{u \dagger} \)
differ only in the singular values
\( x_j \ge 0 \) and \( y_j \ge 0 \) of
the diagonal \(3 \times 3 \) matrices
\( \Sigma^u_h \) and \( \Sigma^u_p  \)\@.
If the mirror fermions have masses
of 1 TeV, then all the Yukawa coupling
constants \(x \) and \( y \) are
between 3 and 5\@.
The full action is invariant under two global
\( U(1) \otimes U(1) \)
symmetries which block mass terms like
\( \bar u_i u_i \) and so forth.
\par
The model implies the existence
of heavy mirror fermions with interactions
much like those of the known fermions,
but with right-handed weak interactions.
They would form very heavy positive
nuclei surrounded by shells
of heavy mirror electrons with
\( m_{e'} \ge 100.8 \) GeV\@.
These mirror atoms would be
very small with Bohr radii
less than \( (\alpha m_{e'})^{-1} \sim 0.27 \) fm---and
less than 0.027 fm
if \( m'_e \ge 1 \) TeV\@.
The energy needed to excite
these atoms would be of the order
of \( m_{e'} \alpha^2 \),
so these atoms
would interact only with photons
of at least an MeV\@.
These atoms are candidates for dark matter.
Because their masses would exceed a TeV,
their number density would be 100 times
lower than that of a 10 GeV \textsc{wimp}\@.
This low number density may be why physicists 
have not detected dark matter even though
its mass density is 5.4 times greater
than that of ordinary matter~\cite{Ade:2015xua}\@.
The quarks of these putative dark-matter particles
interact via QCD, but their
interactions are of very short range
because of the high masses of the exchanged 
heavy pions.
Theories in which dark matter consists of
stealthy strongly interacting 
particles~\cite{Appelquist:2015zfa, *Appelquist:2015yfa} 
or strongly interacting massive particles 
(\textsc{simp}s)~\cite{Hochberg:2014kqa, *Hochberg:2015vrg}
have been developed\@.
Strongly interacting dark matter~\cite{Boddy:2014yra} 
broadens dark-matter cusps into 
cores~\cite{Rocha:2012jg, *Vogelsberger:2012ku} 
as suggested by some observations of galaxies in 
clusters~\cite{Newman:2012nw}
and of stars in nearby galaxies~\cite{Oh:2010mc} 
and so may explain the apparent paucity of heavy
dwarf galaxies around the Milky Way~\cite{BoylanKolchin:2011de, *Zavala:2012us}\@.
\par
The mirror-fermion trick that makes theories
that conserve parity
appear chiral at low energies  
was invented by physicists trying to define
chiral gauge theories on the 
lattice~\cite{Montvay:1987ys,%
*Montvay:1988av, *Csikor:1989km, *Lin:1992qb,% 
*Csikor:1994jg, *Maalampi:1982ak,% 
*Maalampi:1988va, *PhysRevD.50.2297,% 
*Dugan:1991ck, *Dugan:1992gm}\@.
They doubled the number of fermion fields.
A primary set of fermions \( \psi = (\psi_\ell; \psi_r ) \) transforms 
under \( G \supseteq SU_c(3) \otimes SU(2) \otimes U(1) \),
and a secondary set of fermions 
\( \psi' = (\psi'_\ell; \psi'_r ) \) transforms under
\( G' \supseteq SU_c(3) \otimes U(1) \)\@.
When certain spinless fields
assume suitable mean values in the vacuum,
the light fermions are \( \psi_m \simeq (\psi_\ell ; \psi'_r) \)
and have left-handed charged-current weak interactions,
while the heavy fermions are
\( \psi_M \simeq (\psi'_\ell ; \psi_r) \) and
have right-handed charged-current weak interactions.
At low energies the theory looks chiral.
Many physicists have used this mirror-fermion trick.
Anber, Aydemir, Donohue, and 
Pais~\cite{PhysRevD.80.015012} used it to
make a model in which the fermions have vector-like
gauge interactions but chiral Yukawa interactions.
Other physicists~\cite{JHEP09.130, *Raby:2007hm, *Dobrescu:1997nm, *Chivukula:1998wd, *He:1999vp, *Gopalakrishna:2013hua, *Contino:2006qr, *Anastasiou:2009rv, *Vignaroli:2012sf, *DeSimone:2012fs, *Delaunay:2013iia, *Gillioz:2013pba, *Han:2003wu, *Carena:2006jx, *Matsumoto:2008fq, *Berger:2012ec, *Kang:2007ib, *Martin:2009bg, *Graham:2009gy, *Martin:2010dc, *Martin:2012dg, *Fischler:2013tva, *Endo:2011xq, *Endo:2012cc}
have used the trick 
to add vector-like fermions to the 
standard model in various ways, but few have
discussed the spontaneous breaking of parity, 
perhaps because Vafa and Witten
showed~\cite{PhysRevLett.53.535}
that ``in parity-conserving vector-like theories
such as QCD, parity conservation is not
spontaneously broken,'' 
a no-go theorem that does not, however, 
apply to theories with Yukawa 
interactions~\cite{PhysRevLett.53.535, Vafa:1983tf}\@.
Among the few are
Aoki and Gocksch
who exhibited the spontaneous breakdown
of parity in lattice simulations with Wilson 
fermions~\cite{PhysRevLett.57.3136, *Aoki:1987us, *PhysRevD.45.3845}. 
\par
In models with extended gauge groups, 
nucleons slowly decay in processes such as
\( p \to \pi^+ + 3 \nu \),  \( p \to e^+ + 4 \nu \),
and \( n \to 3 \nu \) that 
involve the exchange of three heavy gauge bosons.
These decays 
conserve fermion number
\( F = Q + L \) but violate baryon number
minus lepton number \( B - L \)\@.
Their partial lifetimes 
rise with the 12th power 
of the heavy mass scale \(M\) 
of the mediating gauge bosons
\( \tau_n \sim M^{12} / (\alpha^6_u m_p^{13}) \)
in which \( \alpha_u \) is the 
fine-structure constant of 
the unified theory\@.
The lower bounds on such partial lifetimes
are \(4.9 \times10^{26} \) years for 
\( n \to 3 \nu \)~\cite{Suzuki:1993zp},
\( 5.8 \times10^{29} \) years for 
\( n \to \) invisible~\cite{Araki:2005jt},
and \( 2.1 \times10^{29} \) years
for \( p \to \) invisible~\cite{PhysRevLett.92.102004},
so the masses of the mediating gauge bosons
should exceed about a PeV\@.
Such nucleon-decay events may lurk in SNO, KamLAND, 
Super-Kamiokande, and JUNO data.
The residual excited nucleus \( {}^{15}\)N\(^*\)
or \( {}^{15}\)O\(^*\) emits
a \(\gamma\) ray of 6--7 
MeV~\cite{PhysRevLett.92.102004}\@.
The PeV energy scale is 9 or 10
orders of magnitude lower
than that of traditional grand unification.

\par
Section~\ref{Spin-one-half fields in general relativity}
outlines how Utiyama and Kibble made
theories with fermions compatible with 
general relativity. 
Section~\ref {How to make a vector theory look chiral}
shows how a pseudoscalar field can make
a vector theory look chiral at low energies.
Section~\ref{A model for one generation} 
describes a model for a single generation
of quarks and leptons.
A model for three generations is described
in section~\ref{A model for three generations}\@.
Models of grand unification
with extended gauge groups
are briefly sketched in
Section~\ref{Models with extended gauge groups}\@.
The paper ends with a summary in section~\ref {Summary}\@.

\section{Spin-one-half fields in general relativity
\label {Spin-one-half fields in general relativity}}

Decades ago, Utiyama~\cite{Utiyama1956}
and Kibble~\cite{Kibble1961} showed how
to fit spin-one-half fields into general relativity.
Suppose the flat-space action density is
\begin{equation}
L = - \overline{ \psi } \left[
\gamma^a \left( \partial_a + i g A_a \right) 
+ m \right] \psi
\label {flat spin-one-half action}
\end{equation}
in which \( a \) is a flat-space index,
\( A \) is a matrix of gauge fields,
\( \psi \) is a four-component Dirac or
Majorana field,
\( \overline{ \psi } = \psi^\dagger \beta 
= i \psi^\dagger \gamma^0 \),
and \( m \) is a constant 
or a mean value of a scalar field.
One first introduces tetrad fields
\( e^{\,\,\, \mu}_a(x) \) that turn flat-space
indices \( \gamma^a \) into curved-space indices
\( \gamma^a e^{\,\,\, \mu}_a \)\@.
Derivatives and gauge fields
intrinsically are generally covariant vectors.
So the first step is to replace
\( \gamma^a \left( \partial_a + i g A_a \right) \) by 
\( \gamma^a \, e^{\,\,\, \mu}_a 
\left( \partial_\mu + i g A_\mu \right) \)\@.
The next step is to correct for the effect
of the derivative on the field \( \psi \)
by making the derivative generally covariant
as well as gauge covariant.
The required Einstein connection is
\begin{equation}
E_\mu = \frac{1}{2} \sigma^{a b} \, e_a^{\,\,\, \nu}
\, e_{b \nu ; \mu} 
\label {Lorentz connection}
\end{equation}
in which the \( \sigma^{a b} \)
are the \( 4 \times4 \) matrices
\begin{equation}
\sigma^{a b} = \frac{1}{4} 
\left[ \gamma^a, \gamma^b \right] .
\label {sigmaab}
\end{equation}
The covariant derivative \( e_{b \nu ; \mu} \)
of the tetrad is 
\begin{equation}
e_{b \nu ; \mu} = e_{b \nu , \mu}
- e_{b \sigma} \, \Gamma^\sigma_{\nu \mu}
\label {covariant derivative of a tetrad}
\end{equation}
in which the comma denotes 
and ordinary derivatives, and
\(
\Gamma^\sigma_{\nu \mu} = 
e^\sigma_{\,\,\, a} e^a_{\,\,\, \mu , \nu}
\)
is the Levi-Civita affine connection.
The resulting action density~\cite{Weinberg197212.5} 
\begin{equation}
\mathcal{L} = {} - \overline{ \psi }
\left[ \gamma^a \, e^{\,\,\, \mu}_a
\left( \partial_\mu + i g A_\mu + E_\mu
%+ \frac{1}{2} \sigma^{b \, c} \, e_b^{\,\,\, \nu} \, e_{c \, \nu ; \mu} 
\right) \right] \psi 
\label {L'}
\end{equation}
is invariant \( \mathcal{L}(x) \to \mathcal{L}(x') \)
under any coordinate transformation
\( x \to x' \) that is one-to-one and
differentiable.  
In particular, it is invariant
under the parity transformation 
\( Px = x' = (x^0, - \boldsymbol{x}) \)
which takes the field \( \psi(x) \) to
\begin{equation}
\mathsf{P} \, \psi(x) \, \mathsf{P}^{-1} 
= \eta^* \beta \psi(Px) 
\label{parity transformation}
\end{equation}
in which \( \eta \) is the intrinsic parity
of the particle~\cite{Weinberg19955:5}\@.
\par
One may use this Utiyama-Kibble recipe 
and the usual covariant derivatives of general relativity
to make the action of any of the models of this paper invariant
under general coordinate transformations.

\section {How to make a vector theory look chiral
\label {How to make a vector theory look chiral}}

Most applications of the 
mirror-fermion trick use two or more
scalar fields~\cite{Montvay:1987ys,%
*Montvay:1988av, *Csikor:1989km, *Lin:1992qb,% 
*Csikor:1994jg, *Maalampi:1982ak,% 
*Maalampi:1988va, *PhysRevD.50.2297,% 
*Dugan:1991ck, *Dugan:1992gm, PhysRevD.80.015012,
JHEP09.130, *Raby:2007hm, *Dobrescu:1997nm, *Chivukula:1998wd, *He:1999vp, *Gopalakrishna:2013hua, *Contino:2006qr, *Anastasiou:2009rv, *Vignaroli:2012sf, *DeSimone:2012fs, *Delaunay:2013iia, *Gillioz:2013pba, *Han:2003wu, *Carena:2006jx, *Matsumoto:2008fq, *Berger:2012ec, *Kang:2007ib, *Martin:2009bg, *Graham:2009gy, *Martin:2010dc, *Martin:2012dg, *Fischler:2013tva, *Endo:2011xq, *Endo:2012cc}
and have Yukawa interactions
that explicitly break parity.
The models of this paper use the Higgs
scalar field and one pseudoscalar field 
and have actions that conserve parity.
\par
In this section and the next one,
the scalar field \( h \) and a pseudoscalar field \( p \)
are \( SU(2) \otimes U_Y(1) \) doublets
\begin{equation}
h = \begin{pmatrix} 
      h^+ \\
      h^0 \\
   \end{pmatrix}
   \quad \mbox{and} \quad
p = \begin{pmatrix} 
      p^+ \\
      p^0 \\
   \end{pmatrix}   
   \label {h and p}
\end{equation}
in which \( h^0 = 
( \bar h + h_r + i h_i )/\sqrt{2} \)
and  \( p^0 = ( \bar p + p_r + i p_i )/\sqrt{2} \)\@.
To keep things simple,
we first consider an \( SU_w(2) \otimes U_Y(1) \) 
doublet \( q = (u, d) \) of primary quarks of a given color
and two secondary quarks
\( u' \) and \( d' \) of the same color
that are singlets under \( SU_w(2) \)
but transform under \( U_Y(1) \)
\begin{equation}
q =    \begin{pmatrix} 
      u \\
      d \\
   \end{pmatrix}, \quad
   u', \quad d' .
   \label {udu'd'}
\end{equation}
All the spin-one-half fields of this paper
are four-component Dirac fields.
In the models of this section
and the next,
the Yukawa interactions 
of the primary quarks \( q = (u, d) \) 
and the secondary down quark \( d' \) 
with \( h \) and \( p \) are
\begin{equation}
   \begin{split}
V_d ={}&  \bar q \, ( x_d h + y_d \gamma^5 p ) \, d' 
+  \bar d' \, 
( x_d^* h^\dagger - y_d^* \gamma^5 p^\dagger )\, q 
\\
={}&
u^\dagger_r \, ( x_d h^+ + y_d \gamma^5 p^+ ) \, d'_\ell  
+ u^\dagger_\ell \, ( x_d h^+ + y_d \gamma^5 p^+ ) \, d'_r
\\
+{}& 
\, d^\dagger_r \, ( x_d h^0 + y_d \gamma^5 p^0 ) \, d'_\ell  
+ d^\dagger_\ell \, ( x_d h^0 + y_d \gamma^5 p^0 ) \, d'_r
\\
+{}&
\, d'^\dagger_r \, ( x_d^* h^- - y_d^* \gamma^5 p^- )\, u_\ell
+ d'^\dagger_\ell \, ( x_d^* h^- - y_d^* \gamma^5 p^- )\, u_r
\\
+{}&
\, d'^\dagger_r \, ( x_d^* h^{0 \dagger} 
- y_d^* \gamma^5 p^{0 \dagger} )\, d_\ell
+ d'^\dagger_\ell \, ( x_d^* h^{0 \dagger} 
- y_d^* \gamma^5 p^{0 \dagger} )\, d_r .
   \end{split}
\label {h + gamma5 p}
\end{equation}
They are invariant under space reflection.
Here \( d^\dagger = ( d^\dagger_\ell, d^\dagger_r ) \)
and \( \overline{d} = i d^\dagger \gamma^0 = d^\dagger \beta 
= ( d^\dagger_r, d^\dagger_\ell ) \)\@.
The Dirac matrices obey
\( \{ \gamma^a, \gamma^b \} =  2 \eta^{ab} \)
with \( \eta^{00} = - 1 \) and 
\( \gamma_5 = \gamma^5 = ( 1, 0 ; 0, -1 ) \)\@.
\par
The Yukawa interaction (\ref {h + gamma5 p}) also respects
both the \( U(1) \otimes U(1) \) symmetry
\begin{equation}
   \begin{split}
   q \to {}& e^{i \theta} \gamma^5 q  \quad \mbox{and} \quad
  d' \to  e^{i (\theta - \chi)}  d'  \\
   h \to {}& - e^{i \chi } p \quad \mbox{and} \quad
   p \to  - e^{i \chi } h 
       \label {symmetry 1}
      \end{split}
\end{equation}
which leaves \( \bar q \slashed{D} q \),
\( \bar d' \slashed{D} d' \),
and \( \bar d' d' \)  
but not \( \bar q q \) invariant and
the \( U(1) \otimes U(1) \) symmetry
\begin{equation}
   \begin{split}
    q \to {}& e^{i \theta}  q \quad \mbox{and} \quad
 d' \to  e^{i (\theta - \chi)} \gamma^5 d' \\
   h \to {}& e^{i \chi } p \quad \mbox{and} \quad
   p \to  e^{i \chi } h 
    \label {symmetry 2}
      \end{split}
\end{equation}
which leaves \( \bar q \slashed{D} q \),
\( \bar d' \slashed{D} d' \),
and \( \bar q q \) 
but not \( \bar d' d' \) invariant.
These global symmetries protect
fermion masses by keeping mass terms
like \( m  \bar q q \) and \( m'  \bar d' d' \)
out of the action.
\par
If we replace the fields \( h \) and \( p \) in the
Yukawa interaction (\ref {h + gamma5 p})
by their mean values \( (0, h_0) \) and \( (0, p_0) \) 
in the vacuum, where
\( h_0 = \langle 0 | h^0 | 0 \rangle  = \bar h/\sqrt{2} \)
and \( p_0 = \langle 0 | p^0 | 0 \rangle = \bar p/\sqrt{2} \),
then the Yukawa interaction (\ref {h + gamma5 p}) 
yields the mass terms
\begin{equation}
   \begin{split}
V_{d 0} = {}& \bar d \, ( x_d h_0 + \gamma^5 y_d p_0 ) \, d' 
+ \bar d' \, 
( x^*_0 h^*_0 - \gamma^5 y^*_0 p^*_0 )\, d    \\
={}& d^\dagger_r (x_d h_0 + y_d p_0) d'_\ell
+ d^\dagger_\ell (x_0 h_0 - y_0 p_0 ) d'_r
+ d'^\dagger_r (x^*_0 h^*_0 - y^*_0 p^*_0 ) d_\ell
+ d'^\dagger_\ell (x^*_0 h^*_0 + y^*_0 p^*_0) d_r \\
={}&    \begin{pmatrix} 
      d^\dagger_\ell & d'^\dagger_\ell  
      \end{pmatrix}
  \begin{pmatrix} 
      x_d h_0 - y_d p_0 & 0 \\
   0 & x_d^* h^*_0 + y_d^* p^*_0 \\
   \end{pmatrix}
   \begin{pmatrix} 
       d'_r \\
       d_r    \\
   \end{pmatrix}  + \> \mbox{h.c.}    
\label {LV0}
   \end{split}
\end{equation}
\par
The main self-interactions of the spinless bosons are 
\begin{equation}
V(h,p) = \lambda_h 
\left( h^\dagger h - \frac{v^2}{4} \right)^2
+ \lambda_p
\left( p^\dagger p - \frac{v^2}{4} \right)^2
- \lambda_{h p} 
\left( h^\dagger p + p^\dagger h \right)^2
\label {V(h,p)}
\end{equation}
where \( v= v_{\mathrm{sm}} = 246 \) GeV
and \( \lambda_h = 0.129 \)\@.
The third term in \( V \) has \( 0 < \lambda_{hp} \ll 1 \)
so that the mean values
of \( h^0 \) and \( p^0 \) 
have the same phase (taken to be zero)
or the opposite phase
in which case the light fermions 
would have right-handed weak interactions.
If \( V(h,p) \) were the exact potential of the spinless
fields, the mean values of the neutral
components of the doublets
\( h \) and \( p \) would be
\begin{equation}
h_0 = p_0 = \frac{v}{2}
= 1 2 3 \> \mbox{GeV}.
\label {mean values of h and p}
\end{equation}
For simplicity, I will 
assume that this is the case
and will set
\begin{equation}
| h_0 |^2 + | p_0 |^2 = \frac{v^2}{2}
= \frac{(246)^2}{2} \> \mbox{GeV}^2 .
\label {sum of their squares}
\end{equation}
The field 
\begin{equation}
d_{\mbox{\scriptsize light}} = 
\begin{pmatrix} 
      d_\ell \\
      d'_r \\
   \end{pmatrix}
   \label {d_light}
\end{equation}
then has left-handed \( SU(2) \)
interactions
and a light mass 
\( m_d = | x_d h_0 - y_d p_0 | 
= 123 \, |  x_d  - y_d | \) GeV\@.
Similarly, the field
\begin{equation}
d_{\mbox{\scriptsize heavy}} = 
\begin{pmatrix} 
      d'_\ell \\
      d_r \\
   \end{pmatrix}
   \label {d_light}
\end{equation}
has right-handed \( SU(2) \)
interactions
and a larger mass 
\( M_d = | x_d h_0 + y_d p_0 | 
= 123 \, |x_d + y_d| \) GeV,
which can be
as heavy as 3 TeV with
\( |x_d | \le 4 \pi \) and \( | y_d | \le 4 \pi \)\@.
Thus a theory whose action conserves parity 
can look chiral at low energies.
If we take \( y_d \) and \( z_d \) 
to be positive, then the
Yukawa coefficients are
\begin{equation}
x_d = \frac{M_d + m_d}{v}
\quad \mbox{and} \quad
y_d = \frac{M_d - m_d}{v} .
\label {formula for y and z}
\end{equation}
Since \( M_d \gtrsim 700 \) GeV
while \( m_d = 4.8 \) MeV,
these Yukawa coefficients are nearly equal,
\( x_d \approx y_d \gtrsim 2.8 \),
because \( M_d \gg m_d \) 
and because I set \( h_0 = p_0 \)\@.
\par
Before passing to
a more complete model,
let's note what happens
when a scalar field \( s \)
replaces the pseudoscalar field \( p \)\@. 
The minus sign 
in our Yukawa interaction (\ref {h + gamma5 p})
arose because \( \gamma^0 \)
and \( \gamma^5 \) anticommute,
so the minus sign goes away, 
and instead of \( V_d \) we have
\begin{equation}
V_{d s} = \bar q \, ( x_d h + y_d s ) \, d' 
+ \bar d' \, 
( x^*_d h^\dagger + y^*_d s^\dagger )\, q 
\label {h + s}
\end{equation}
in which \( x_d \)
and \( y_d \) are coupling constants.
If the neutral components of the
scalar fields \( h \) and \( s \)
assume the mean values \( h_0 \) and \( s_0 \)
in the vacuum, then the mass terms 
of \( V_{d s} \) are
\begin{equation}
V_{ds0} = {} \begin{pmatrix} 
      d^\dagger_\ell & d'^\dagger_\ell  
      \end{pmatrix}
  \begin{pmatrix} 
      x_d h_0  + y_d p_0 & 0 \\
   0 & x^*_d h^*_0 + y^*_d p^*_0 \\
   \end{pmatrix}
   \begin{pmatrix} 
       d'_r \\
       d_r    \\
   \end{pmatrix}  + \> \mbox{h.c.} 
\label {the h-s mass terms are}
\end{equation}
The vacuum values \( h_0 \) and \( s_0 \)
give the same mass to 
\( (d_\ell, d'_r) \) and \( (d'_\ell, d_r) \)
and so do not break parity spontaneously.

\section{A model for one generation
\label {A model for one generation}}

In the proposed model,
the vacuum breaks parity by giving 
a mean value to a new pseudoscalar field \( p \)
which makes the mirror-fermion trick of
section~\ref {How to make a vector theory look chiral} work.
The action of the model is invariant
under general coordinate transformations
when suitably decorated with tetrads
as in section~\ref {Spin-one-half fields in general relativity}\@.
The model has a secondary fermion
for each (primary) fermion of the standard model.
A gauge group 
\( G \supseteq SU_c(3) \otimes SU_w(2) \otimes U_Y(1) \)
acts on the four-component primary fermions,
and a group
 \( G' \supseteq SU_c(3) \otimes U_Y(1) \)
acts on the secondary fermions.

\begin{table}[h]
\begin{center}
\begin{tabular}{l c c c c c c c c}
\hline\hline
\quad & \quad 
\( q =    \begin{pmatrix} 
      u \\
      d \\
   \end{pmatrix} \) \quad &
\quad   \( \ell =    \begin{pmatrix} 
      \nu \\
      e \\
   \end{pmatrix} \) \quad &
\quad \( u' \) \quad & \quad \(  d'  \) \quad &
\quad \( \nu'  \) \quad &  \quad \( e' \) \quad &
 \quad  \( h =    \begin{pmatrix} 
      h^+ \\
      h^0 \\
   \end{pmatrix} \) \quad & 
   \quad \( p  =    \begin{pmatrix} 
      p^+ \\
      p^0 \\
   \end{pmatrix} \) \\
  isospin \( t \) & \(2\) & {} \(2\) & 1 & 1 & 1 & 
   {} 1 & 2 & 2 \\
  hypercharge \( y \) & \(\frac{1}{6}\) &  \( - \frac{1}{2}\) & \(\frac{2}{3}\) & \(-\frac{1}{3}\) & 0 & 
   {}\( -1\) & \(\frac{1}{2}\) & \(\frac{1}{2}\) \\
   color \( c \) & 3 & 1 & 3 & 3 & 1 & 
   {} 1 & 1 & 1 \\
   \hline\hline
\end{tabular}
\end{center}
\caption{One generation of fermions and the scalar \(h\)
and pseudoscalar \(p\) fields of the simplest model.
The integers \( t \) and \( c \) are the dimension
of the representation of \( SU_w(2) \) and 
of \( SU_c(3) \)\@.}
\label{simplest model}
\end{table}

In the model,
the pseudoscalar field \( p\)
is a doublet that transforms under
\( SU_w(2) \otimes U_Y(1) \) like
the Higgs doublet.  
So for \( g \in SU_w(2) \otimes U_Y(1) \)
and \( g' \in U_Y(1) \),
the fields transform as
\( q \to g q \), \( \ell \to g \ell \), 
\( h \to g h \), \( p \to g p \),
\( u' \to g' u' \), \( d' \to g' d' \),
\( \nu' \to g' \nu' \), and \( e' \to g' e' \)\@.
The covariant derivatives of
the primary quark 
and lepton doublets of the first generation are
\begin{equation}
   \begin{split}
   D_\mu q ={}& \left( \partial_\mu + i g T_a A^a_\mu
+ i g' V B_\mu \right) q
={} \left( \partial_\mu + i g \frac{\sigma_a}{2} A^a_\mu
+ i g' \frac{1}{6} B_\mu  \right) q \\
D_\mu \ell ={}& \left( \partial_\mu + i g T_a A^a_\mu
+ i g' V B_\mu \right) \ell
={} \left( \partial_\mu + i g \frac{\sigma_a}{2} A^a_\mu
- i g' \frac{1}{2} B_\mu  \right) \ell ,
\label {covariant derivative acting on the primary quarks and leptons}
   \end{split}
\end{equation}
while those of the secondary quarks 
\( u' \) and \( d' \) and 
leptons \( \nu' \) and \( e' \) are
\begin{equation}
   \begin{split}
   D_\mu u' ={}& \left( \partial_\mu 
+ i g' V B_\mu \right) u'
={} \left( \partial_\mu 
+ i g'  \frac{2}{3} B_\mu \right) u' \\
   D_\mu d' ={}& \left( \partial_\mu 
+ i g' V B_\mu \right) d' 
={} \left( \partial_\mu 
- i g' \frac{1}{3} B_\mu \right) d'  \\
   D_\mu \nu' ={}& \left( \partial_\mu 
+ i g' V B_\mu \right) \nu'
={}  \partial_\mu \, \nu' \\
   D_\mu e' ={}& \left( \partial_\mu 
+ i g' V B_\mu \right) e'
={} \left( \partial_\mu 
- i g' B_\mu \right) e' .
\label {covariant derivatives acting on the secondary quarks and leptons}
   \end{split}
\end{equation}
The first-generation Yukawa terms are
\begin{equation}
   \begin{split}
 V ={}&   \bar q \, i \tau_2 
   ( x_u h^* + y_u\gamma^5 p^*)u' 
 -  \bar u'  
 ( x_u^*h^\intercal - y_u^* \gamma^5 p^\intercal) \, i \tau_2 q  
 \\
 {}& +   \bar q ( x_d h + y_d \gamma^5 p )d' 
 +  \bar d'  ( x_d^* h^\dagger  - y_d^* \gamma^5 p^\dagger ) q 
 \\
 {}& +   \bar \ell \, i \tau_2 
 ( x_\nu h^* + y_\nu \gamma^5 p^* )  \nu' 
 -  \bar \nu' 
 ( x_\nu^* h^\intercal - y_\nu^* \gamma^5 p^\intercal ) 
 \, i\tau_2 \ell  
 \\
 {}& +   \bar \ell ( x_e h + y_e \gamma^5 p ) e' 
 +  \bar e' ( x_e^* h^\dagger - y_e^* \gamma^5 p^\dagger ) \ell 
    \label {LVs p doublet}
   \end{split}
\end{equation}
in which \( \tau_2 \) is the second Pauli matrix.
The Yukawa terms (\ref {LVs p doublet}) and 
the kinetic terms
\( \bar q \slashed{D} q \), \( \bar \ell \slashed{D} \ell \),
\( \bar u' \slashed{D} u' \), \( \bar d' \slashed{D} d' \), 
\( \bar \nu' \slashed{D} \nu' \), and \( \bar e' \slashed{D} e' \)
in the action of the model are
invariant under the gauged
symmetry \( G \otimes G' \) and
under the two global
\( U(1) \otimes U(1) \) symmetries 
\begin{equation}
   \begin{split}
   q \to {}& e^{i \theta} \gamma^5 q  \quad \mbox{and} \quad
\ell \to  e^{i \theta} \gamma^5 \ell  \\
   u' \to {}& e^{i (\theta + \chi)} u'  \quad \mbox{and} \quad
   \nu' \to e^{i (\theta + \chi)} \nu'   \\
  d' \to {}& - e^{i (\theta - \chi)}  d'  \quad \mbox{and} \quad
e' \to  - e^{i (\theta - \chi)}  e'   \\   
   h \to {}& - e^{i \chi } p \quad \mbox{and} \quad
   p \to  - e^{i \chi } h 
       \label {symmetry 1}
      \end{split}
\end{equation}
and
\begin{equation}
   \begin{split}
    q \to {}& e^{i \theta}  q \quad \mbox{and} \quad
\ell \to  e^{i \theta}  \ell \\
   u' \to {}& e^{i (\theta + \chi)} 
   \gamma^5 u' \quad \mbox{and} \quad
   \nu' \to  e^{i (\theta + \chi)} \gamma^5 \nu'  \\
 d' \to {}& e^{i (\theta - \chi)} \gamma^5 d' 
 \quad \mbox{and} \quad
e' \to  e^{i (\theta - \chi)} \gamma^5 e'   \\   
   h \to {}& e^{i \chi } p \quad \mbox{and} \quad
   p \to  e^{i \chi } h .
    \label {symmetry 2}
      \end{split}
\end{equation}
But the global symmetry (\ref {symmetry 1})
changes \( \bar u u \), \dots, \( \bar e e \),
and the other global symmetry (\ref {symmetry 2})
changes \( \bar u' u' \), \dots, \( \bar e' e' \)\@.
So these global symmetries 
keep mass terms like \( m_u \bar u u \),
\( m_{u'} \bar u' u' \),
and so forth out of an action
that consists only of the Yukawa 
interaction (\ref {LVs p doublet}),
the kinetic terms \( \bar q \slashed{D} q \),
and so forth.
\par
Replacing the spinless bosons in 
the Yukawa interaction
(\ref {LVs p doublet}) by their
mean values in the vacuum
and grouping fields of the same
handedness together,
we get 
\begin{equation}
   \begin{split}
V_0 ={}& \begin{pmatrix} 
    u^\dagger_\ell   &  
    u'^\dagger_\ell  
   \end{pmatrix}
   \begin{pmatrix} 
 x^*_u h^*_0 - y^*_u p^*_0   &  0 \\
  0 & x^*_u  h_0 + y^*_u p_0    \\
    \end{pmatrix}
     \begin{pmatrix} 
    u'_r \\
    u_r   
    \end{pmatrix}  
    \\
{}& +
\begin{pmatrix} 
    d^\dagger_\ell   &  
    d'^\dagger_\ell  
   \end{pmatrix}
   \begin{pmatrix} 
 x_d h_0 - y_d p_0  &  0 \\
 0 &  x^*_d  h^*_0 + y^*_d p^*_0 ) \\
    \end{pmatrix}
     \begin{pmatrix} 
    d'_r \\
    d_r   
    \end{pmatrix}  
    \\
{}&    +
\begin{pmatrix} 
    \nu^\dagger_\ell   &  
    \nu'^\dagger_\ell  
   \end{pmatrix}
   \begin{pmatrix} 
 x_{\nu_e} h^*_0 - y^*_{\nu_e} p^*_0 )  &  0 \\
  0 & x^*_{\nu_e} h_0 + y^*_{\nu_e} p_0 )   \\
    \end{pmatrix} 
     \begin{pmatrix} 
    \nu'_r \\
    \nu_r   
    \end{pmatrix}  
    \\
{}&    +
    \begin{pmatrix} 
    e^\dagger_\ell   &  
    e'^\dagger_\ell  
   \end{pmatrix}
   \begin{pmatrix} 
 x_e  h_0 -  y_e p_0 )  &  0 \\
 0 & x^*_e h^*_0 + y^*_e p^*_0 )  \\
    \end{pmatrix}
     \begin{pmatrix} 
    e'_r \\
    e_r   
    \end{pmatrix}   + \> \mbox{h.c.}  
           \end{split}
           \label {generation matrix}
\end{equation}
We set 
\begin{equation}
|h_0|^2 + |p_0|^2 = {\textstyle{ \frac{1}{2} }}  v^2%_{\mathrm{sm}} 
= {\textstyle{ \frac{1}{2} }}  (246 \,\, \mbox{GeV})^2
\label {vsm}
\end{equation}
and keep \( h_0 = p_0 = 123 \) GeV 
as in (\ref {mean values of h and p})\@.
The light-mass fields of the first generation are
\begin{equation}
u_m =  \begin{pmatrix}
u_\ell \\
u'_r
\end{pmatrix}, \quad
d_m =  \begin{pmatrix}
d_\ell \\
d'_r
\end{pmatrix}, \quad
\nu_m =  \begin{pmatrix}
\nu_\ell \\
\nu'_r
\end{pmatrix}, \quad
e_m =  \begin{pmatrix}
e_\ell \\
e'_r
\end{pmatrix} .
\label {light-mass fields of the first generation}
\end{equation}
Their masses are 
\begin{equation}
   \begin{split}  
m_u ={}& | x_u h_0 - y_u p_0 |
\qquad
m_d = | x_d h_0 - y_d p_0 |
\\
m_{\nu_e} ={}& | x_{\nu_e} h_0 - y_{\nu_e} p_0 |
\qquad
m_e = | x_e h_0 - y_e p_0 | .
\label {1 generation light masses}
   \end{split}
\end{equation}
The heavy-mass fields are
\begin{equation}
u_M =  \begin{pmatrix}
u'_\ell \\
u_r
\end{pmatrix}, \quad 
d_M =  \begin{pmatrix}
d'_\ell \\
d_r
\end{pmatrix} , \quad
\nu_M =  \begin{pmatrix}
\nu'_\ell \\
\nu_r
\end{pmatrix}, \quad
e_M =  \begin{pmatrix}
e'_\ell \\
e_r
\end{pmatrix} .
\label {the light and heavy mass eigenstates are}
\end{equation}
Their masses are 
\begin{equation}
   \begin{split}  
M_u ={}& | x_u h_0 + y_u p_0 |
\qquad
M_d = | x_d h_0 + y_d p_0 |
\\
M_{\nu_e} ={}& | x_{\nu_e} h_0 + y_{\nu_e} p_0 |
\qquad
M_e = | x_e h_0 + y_e p_0 | .
\label {1 generation heavy masses}
   \end{split}
\end{equation}
The Yukawa coefficients are
\begin{equation}
x_u = \frac{M_u + m_u}{2 h_0}
\quad \mbox{and} \quad
y_u = \frac{M_u - m_u}{2 p_0} 
\label {The Yukawa coefficients are}
\end{equation}
with similar formulas for \( d  \),
\( \nu_e \), and \( e \)\@.
So we assume the equality
(\ref {mean values of h and p})
of the mean values \( h_0 \)
and \( p_0 \), then the \(x\)'s
and the \(y\)'s are nearly equal
and less than \( 4 \pi \)\@.
For instance, if \( M_u = 750 \) GeV,
then \( x_u = 750/246 = 3.05 \)
and \( y_u \approx x_u \)\@.
\par% edited so far
With \( L = ( 1 + \gamma^5)/2 \)
and \( R = ( 1 - \gamma^5)/2 \),
the light and heavy quark doublets are
\begin{equation}
q_m = L q + R q'
\quad \mbox{and} \quad
q_M = R q + L q' .
\label {light and heavy quark doublet}
\end{equation}
Thus \( L q = L q_m \) and \( Rq = R q_M \),
so that \( q = (L+R) q = L q_m + R q_M \)\@.
The covariant derivative 
\( \left( \partial_\mu + \frac{i g}{2} \sigma_a A^a_\mu
+ \frac{i g'}{6}  B_\mu  \right) q \)
of the primary quarks 
(\ref{covariant derivative acting on the primary quarks and leptons}) therefore acts on 
the left-handed light quarks \( L q_m \)
and on the right-handed heavy quarks \( R q_M \)
\begin{equation}
\bar q \slashed{D}q ={}
\overline {(L q_m + R q_M)} \,
\slashed{D} (L q_m + R q_M)
= \overline {L q_m} \, \slashed{D} L q_m
+ \overline {R q_M} \, \slashed{D} R q_M
\label {kinetic action of the primary quarks}
\end{equation}
which accordingly interact with all the 
\( SU_w(2) \otimes U_Y(1) \) gauge bosons.
An analogous rule applies to the 
primary leptons.
\par
Similarly, \( Rq' = Rq_m \) and 
\( L q' = L q_M \), so that
\( q' = Rq_m + L q_M \)\@.
The covariant derivative 
\( \left( \partial_\mu 
+   \frac{2i g'}{3} B_\mu \right) \)
of the secondary up quark 
(\ref{covariant derivatives acting on the secondary quarks and leptons})
therefore acts on 
the right-handed light up quark \( R u_m \)
and on the left-handed heavy up quark \( L U_M \)
\begin{equation}
\bar u' \slashed{D} u' ={}
\overline {(R u_m + L u_M)} \,
\slashed{D} (R u_m + L u_M)
= \overline {R u_m} \, \slashed{D} R u_m
+ \overline {L u_M} \, \slashed{D} L u_M 
\label {kinetic action of the u' quark}
\end{equation}
which accordingly interacts only with 
the \( U_Y(1) \) gauge boson.
Analogous rules apply to the other
secondary fermions.
\par
To see if we can keep 
the \( W \) and \( Z \) gauge bosons 
and the photon at their physical masses,
we note that the covariant derivative
acting on \( h \) and on \( p \),
which have \( y = \frac{1}{2} \), is
\begin{equation}
   \begin{split}
D_\mu ={}& \partial_\mu + i g T_a A^a_\mu
+ i g' V B_\mu 
={} \partial_\mu + i g \frac{\sigma_a}{2} A^a_\mu
+ i g' \frac{1}{2} B_\mu .
\label {Higgs covariant derivative}
   \end{split}
\end{equation}
The mean value in the vacuum of \( D_\mu h \) is
\begin{equation}
\langle D_\mu h \rangle_0 ={} \left( i g \frac{\sigma_a}{2} A^a_\mu
+ i g' \frac{1}{2} B_\mu \right)    \begin{pmatrix} 
      0 \\
      h_0 \\
   \end{pmatrix}
   ={} i \frac{h_0}{2}
   \begin{pmatrix} 
   g(A^1_\mu -i A^2_\mu) \\
   - g A^3_\mu + g' B_\mu
    \end{pmatrix},
    \label {vacuum covariant derivative of h}
\end{equation}
and that of \( D_\mu p \) is
\begin{equation}
\langle D_\mu p \rangle_0 ={} \left( i g \frac{\sigma_a}{2} A^a_\mu
+ i g' \frac{1}{2} B_\mu \right)    \begin{pmatrix} 
      0 \\
      p_0 \\
   \end{pmatrix}
   ={} i \frac{p_0}{2}
   \begin{pmatrix} 
   g(A^1_\mu -i A^2_\mu) \\
   - g A^3_\mu + g' B_\mu
    \end{pmatrix}.
    \label {vacuum covariant derivative of p}
\end{equation}
The mass terms of the gauge bosons
are
\begin{equation}
\begin{split}
\langle \, (D_\mu h)^\dagger D^\mu h \, \rangle_0 +
\langle \, (D_\mu p)^\dagger D^\mu p \, \rangle_0 ={}&
\frac{|h_0|^2 + |p_0|^2}{4}
\Big[ g^2 \left( A^1_\mu  A^{1 \mu}+ A^2_\mu A^{2 _\mu} \right) \\
{}& + \left( -g A^3_\mu + g' B_\mu \right) 
\left( -g A^{3 \mu} + g' B^\mu \right) \Big] .
\label {mass terms of the gauge bosons}
\end{split}
\end{equation}
The photon \( g' A^3_\mu + g B_\mu \)
is massless, and the 
\( W^\pm \) and \( Z \)
\begin{equation}
   \begin{split}
W^\pm_\mu = {}&
\frac{ A^1_\mu \mp i A^2_\mu }{\sqrt{2}} \quad
\mbox{and} \quad
Z_\mu  =  \frac{g A^3_\mu - g' B_\mu}
{\sqrt{g^2 + g'^2 }} \\
   \end{split}
\end{equation}
get their physical masses
\begin{equation}
   \begin{split}
      m_W = {}&  \frac{g \, v}{2} \quad
      \mbox{and} \quad
      m_Z =  \frac{ \sqrt{g^2 + g'^2} \, v}
      {2} .
   \end{split}
\end{equation}
So the electro-weak gauge bosons
have their usual masses as long as
the squares of their mean values (\ref {vsm})
add up to \( h_0^2 + p_0^2 = v^2/2 = (246)^2/2 \) GeV\(^2\)\@.
The choice (\ref {mean values of h and p})
of equal mean values 
\( h_0 = p_0  = 123 \) GeV
is a simple way to satisfy this constraint.

\section{A model for three generations
\label {A model for three generations}}

The model for three generations
is essentially three copies of the
model for one generation.
Its Yukawa sector 
has three doublets of primary quarks
\begin{equation}
q_1 =    \begin{pmatrix} 
      u \\
      d \\
   \end{pmatrix}
   \qquad
q_2 =    \begin{pmatrix} 
      c \\
      s \\
   \end{pmatrix}
   \qquad   
q_3 =    \begin{pmatrix} 
      t \\
      b \\
   \end{pmatrix}   
\end{equation}
and three doublets of primary leptons
\begin{equation}
\ell_1 =    \begin{pmatrix} 
      \nu_e \\
      e \\
   \end{pmatrix}
   \qquad
\ell_2 =    \begin{pmatrix} 
      \nu_\mu \\
      \mu \\
   \end{pmatrix}
   \qquad   
\ell_3 =    \begin{pmatrix} 
      \nu_\tau \\
      \tau \\
   \end{pmatrix}   
\end{equation}
as well as secondary quarks and leptons
\(u'_1 = u'\), \(u'_2 = c' \), \(u'_3 = t' \),
\dots \( e'_1 = e \), \( e'_2 = \mu' \),
and \( e'_3 = \tau' \) that are singlets
under \( SU(2) \otimes U(1) \)\@.
\par
The model avoids
flavor-changing neutral currents
by means of a new form of Yukawa alignment
in which the \(3 \times 3 \) matrices
that couple the fields \( h \) and \( p \)
to the fermions have 
singular-value decompositions
that differ only in their singular values.
For instance, the Yukawa matrices 
\( U^u \Sigma^u_h V^{u \dagger} \)
and
\( U^u \Sigma^u_p V^{u \dagger} \)
that give
masses to the three generations
of up quarks
and mirror up quarks 
differ only in the \(3 \times 3 \) diagonal matrices
\( \Sigma^u_h \) and \( \Sigma^u_p  \)
of singular values
\( x_j \ge 0 \) and \( y_j \ge 0 \)\@.
If \( q_i = (u_i, d_i) \) for \( i = 1, 2, 3 \)
are the primary quarks,
\( u'_k \) and \( d'_k \)
the secondary quarks,
\( \ell_i  = (\nu_i, e_i) \) the primary
leptons, and \( \nu'_k \)
and \( e'_k \) the secondary leptons,
then the Yukawa interactions are
\begin{equation}
   \begin{split}
V ={} & \sum_{i, j, k = 1}^3
\bar q_i  \, i \tau_2 \, U^u_{i j} 
\left( x^u_j h + y^u_j \gamma^5 p \right) V^{u \dagger}_{jk} 
\, u'_k
+
\sum_{i, j, k = 1}^3
\bar q_i  \, U^d_{i j} 
\left( x^d_j h + y^d_j \gamma^5 p \right) V^{d \dagger}_{j k} 
\, d'_k 
\\
+ {}&
\sum_{i, j, k = 1}^3
\bar \ell_i  \, i \tau_2 \, U^{\nu}_{i j} 
\left( x^\nu_j h + y^\nu_j \gamma^5 p \right) V^{\nu \dagger}_{jk} 
\, \nu'_k
+
\sum_{i, j, k = 1}^3
\bar \ell_i  \, U^e_{i j} 
\left( x^e_j h + y^e_j \gamma^5 p \right) V^{d \dagger}_{j k} 
\, e'_k + \mbox{h.c.}
   \end{split}
\end{equation}
The left and right singular vectors are
\begin{equation}
\begin{split}
\bar u^s_j = {} & \sum_{i=1}^3 \bar u_i \, U^u_{i j}
\qquad
\bar d^s_j = {} \sum_{i=1}^3 \bar d_i \, U^d_{i j}
\qquad
u'^s_j = \sum_{k=1}^3 V^{u \dagger}_{jk} \, u'_k
\qquad
d'^s_j = \sum_{k=1}^3 V^{d \dagger}_{jk} \, d'_k
\\
\bar \nu^s_j = {} & \sum_{i=1}^3 \bar \nu_i \, U^\nu_{i j}
\qquad
\bar e^s_j = {}  \sum_{i=1}^3 \bar e_i \, U^e_{i j}
\qquad
\nu'^s_j = \sum_{k=1}^3 V^{\nu \dagger}_{jk} \, \nu'_k
\qquad
e'^s_j = \sum_{k=1}^3 V^{e \dagger}_{jk} \, e'_k .
\end{split}
\end{equation}
Replacing the scalar and pseudoscalar doublets
by their neutral components
\( h^0 \) and \( p^0\), we see that
the Yukawa interactions \( V \) 
make no neutral currents
\begin{equation}
   \begin{split}
V ={} & \sum_{j = 1}^3
\bar u^s_j  
\left( x^u_j h^0 + y^u_j \gamma^5 p^0 \right) u'^s_j
+
\sum_{j = 1}^3
\bar d^s_j   
\left( x^d_j h^0 + y^d_j \gamma^5 p^0 \right)  d'^s_j 
\\
+ {}&
\sum_{j = 1}^3
\bar \nu^s_j  
\left( x^\nu_j h^0 + y^\nu_j \gamma^5 p^0 \right) \nu'^s_j
+
\sum_{j = 1}^3
\bar e^s_j  
\left( x^e_j h^0 + y^e_j \gamma^5 p^0 \right) e'^s_j + \mbox{h.c.}
\label {V for 3 generations}
   \end{split}
\end{equation}
The CKM matrices of the quarks and leptons are
\begin{equation}
W^q_{\textsc{ckm}} ={} U^{u \dagger} U^d 
\quad \mbox{and} \quad
W^\ell_{\textsc{ckm}} ={} U^{\nu \dagger} U^e .
\label {CKM matrices of the quarks and leptons}
\end{equation}
If the secondary fermions have their own 
\( SU(2)' \),
then their quark and lepton CKM matrices are
\begin{equation}
W^{q'}_{\textsc{ckm}} ={} V^{u' \dagger} V^{d'}
\quad \mbox{and} \quad
W^{\ell'}_{\textsc{ckm}} ={} V^{\nu' \dagger} V^{e'} .
\label{mirror CKM matrices}
\end{equation}
The CKM matrices \( W^\ell_{\textsc{ckm}} \),
\( W^{q'}_{\textsc{ckm}} \), and
\( W^{\ell'}_{\textsc{ckm}} \)
may break \( CP \) enough to explain
why there's so much more matter than antimatter.
\par
Replacing \( h^0 \) and \( p^0 \) in the 
Yukawa potential (\ref {V for 3 generations})
by their mean values
in the vacuum \( h_0 \) and \( p_0 \), we get
the mass terms
\begin{equation}
   \begin{split}
V_0 ={} & \sum_{j = 1}^3
\bar u^s_j  
\left( x^u_j h_0 + y^u_j \gamma^5 p_0 \right) u'^s_j
+
\sum_{j = 1}^3
\bar d^s_j   
\left( x^d_j h_0 + y^d_j \gamma^5 p_0 \right)  d'^s_j 
\\
+ {}&
\sum_{j = 1}^3
\bar \nu^s_j  
\left( x^\nu_j h_0 + y^\nu_j \gamma^5 p_0 \right) \nu'^s_j
+
\sum_{j = 1}^3
\bar e^s_j  
\left( x^e_j h_0 + y^e_j \gamma^5 p_0 \right) e'^s_j + \mbox{h.c.} 
   \end{split}
\end{equation}
Thus by analogy with the one-generation case
(\ref {light-mass fields of the first generation}--\ref
{The Yukawa coefficients are}), 
the light-mass fields are
\begin{equation}
u_{m j} =  \begin{pmatrix}
u^s_{\ell j}\\
u'^s_{r j}
\end{pmatrix}, \quad
d_{m j} =  \begin{pmatrix}
d^s_{\ell j}\\
d'^s_{r j}
\end{pmatrix}, \quad
\nu_{m j} =  \begin{pmatrix}
\nu^s_{\ell j} \\
\nu'^s_{r j}
\end{pmatrix}, \quad
e_{m j} =  \begin{pmatrix}
e^s_{\ell j}\\
e'^s_{r j}
\end{pmatrix} .
\label {light-mass fields of three generations}
\end{equation}
The singular values \( x^u_j, y^u_j \dots x^e_j, y^e_j \)
are nonnegative, so we take \( h_0 \)
and \( p_0 \) to be positive.
Thus the masses of the light particles are
\begin{equation}
   \begin{split}  
m_{u j} ={}&  x^u_j h_0 - y^u_j p_0 
\qquad
m_{d j} =  x^d_j h_0 - y^d_j p_0 
\\
m_{\nu_e j} ={}&  x^{\nu_e}_j h_0 - y^{\nu_e}_j p_0 
\qquad
m_{e j} =  x^e_j h_0 - y^e_j p_0  .
\label {3 generations light masses}
   \end{split}
\end{equation}
The heavy-mass fields are
\begin{equation}
u_{M j} =  \begin{pmatrix}
u'^s_{\ell j} \\
u^s_{r j}
\end{pmatrix}, \quad 
d_{M j} =  \begin{pmatrix}
d'^s_{\ell j} \\
d^s_{r j}
\end{pmatrix} , \quad
\nu_{M j} =  \begin{pmatrix}
\nu'^s_{\ell j} \\
\nu^s_{r j}
\end{pmatrix}, \quad
e_{M j} =  \begin{pmatrix}
e'^s_{\ell j} \\
e^s_{r j}
\end{pmatrix} .
\label {3 generations heavy masses}
\end{equation}
Their masses are 
\begin{equation}
   \begin{split}  
M_{u j} ={}&  x^u_j h_0 + y^u_j p_0 
\qquad
M_{d j} =  x^d_j h_0 + y^d_j p_0 
\\
M_{\nu_e j} ={}&  x^{\nu_e}_j h_0 + y^{\nu_e}_j p_0 
\qquad
M_{e j} =  x^e_j h_0 + y^e_j p_0  .
\label {3 generation heavy masses}
   \end{split}
\end{equation}
The Yukawa coefficients are
\begin{equation}
\begin{split}  
x^u_j ={}& \frac{M_{u_j} + m_{u_j}}{2 h_0}
\quad 
y^u_j ={} \frac{M_{u_j} - m_{u_j}}{2 p_0}
\qquad
x^d_j ={} \frac{M_{d_j} + m_{d_j}}{2 h_0}
\quad 
y^d_j ={} \frac{M_{d_j} - m_{d_j}}{2 p_0}
\\
x^\nu_j ={}& \frac{M_{\nu_j} + m_{\nu_j}}{2 h_0}
\quad 
y^\nu_j ={} \frac{M_{\nu_j} - m_{\nu_j}}{2 p_0}
\qquad
x^e_j ={} \frac{M_{e_j} + m_{e_j}}{2 h_0}
\quad 
y^e_j ={} \frac{M_{e_j} - m_{e_j}}{2 p_0} .
\end{split}
\end{equation}
\par
The Particle Data Group~\cite{Patrignani:2016xqp} quark masses are
\( m_u = 2.3 \),
\( m_d = 4.8 \), and
\( m_s = 95 \) MeV;
and 
\( m_c = 1.275 \),
\( m_b = 4.66 \), and
\( m_t = 173.1 \) GeV\@.
The PDG masses of the charged leptons are
\( m_e = 0.511 \), 
\( m_\mu = 105.66 \), and
\( m_\tau = 1776.82 \) MeV\@.
The neutrino masses are 
unknown, but the PDG estimates are
\( m_{\nu_e} < 2 \) eV, and
\( m_{\nu_\mu} < 0.19 \) and
\( m_{\nu_\tau} < 18.2\) MeV\@.
The PDG lower limits on the masses
of heavy fermions are 
\(m_t' > 700 \) GeV, \( m_b' > 675 \) GeV,
and \( m_\tau' > 100.8 \) GeV\@.
The lower limits on the mass of a fourth generation
\( t' \) quark run from 350 to 782 
GeV~\cite{Agashe:2014kda}\@.
If we choose \( h_0 = p_0 = 123 \) GeV
and assume that the masses of the heavy
particles are 1 TeV,
then the largest Yukawa coefficient is
\begin{equation}
x^u_3 = {} \frac{M_{u_3} + m_{u_3}}{2 h_0}
= \frac{1173}{246} = 4.77 
\label {biggest Yukawa coefficient}
\end{equation}
which is less than \( 4 \pi \)\@.
The smallest coefficient is 
\begin{equation}
y^u_3 = {} \frac{M_{u_3} - m_{u_3}}{2 p_0}
= \frac{827}{246} = 3.36.
\label {biggest y Yukawa coefficient}
\end{equation}
Under the same assumptions,
\(M_{u 1} = 1 \) TeV and 
\( h_0 = p_0 = 123 \) GeV,
the \( x \) and \( y \) coefficients
of the first generation
are much closer together:
\begin{equation}
x^u_1 = {} \frac{M_{u_1} + m_{u_1}}{2 h_0}
= 4.06505
\quad \mbox{and} \quad
y^u_1 = {} \frac{M_{u_1} - m_{u_1}}{2 p_0}
= 4.06503 .
\label {first generation x y coefficients}
\end{equation}
Inasmuch as the mass \( M_{u_1} \) is unknown,
the extra digits are meant only
to suggest how close \(x^u_1\) is to \(y^u_1\)
and not what their actual values are.
Under the same assumptions,
\(M_{e 1} = M_{e 3} = 1 \) TeV and 
\( h_0 = p_0 = 123 \) GeV,
the coefficients of the charged leptons
of the first generation are
\begin{equation}
x^e_1 = {} \frac{M_{e_1} + m_{e_1}}{2 h_0}
= 4.065043
\quad \mbox{and} \quad
y^e_1 = {} \frac{M_{e_1} - m_{e_1}}{2 p_0}
= 4.065039 
\label {first generation x y coefficients}
\end{equation}
and those of the third generation are
\begin{equation}
x^e_3 = {} \frac{M_{e_3} + m_{e_3}}{2 h_0}
= 4.072
\quad \mbox{and} \quad
y^e_3 = {} \frac{M_{e_3} - m_{e_3}}{2 p_0}
= 4.058 .
\label {first generation x y coefficients}
\end{equation}
These Yukawa coefficients are bigger
than the ones computed from the 
standard model and the observed fermion masses,
but they all are less than \( 4 \pi \), and
the mass mechanism
(\ref {3 generations light masses})
is different from that of the standard model.
\par
The clustering of the Yukawa coefficients
of all three generations
about values near \( 4 \) 
is due to the simplifying assumptions that 
\( h_0 = p_0\) and that
all the heavy masses are 1 TeV\@.
Future measurements
of Higgs decays will determine
whether the Yukawa coefficients
look at all like 
(\ref {biggest Yukawa coefficient}--\ref
{first generation x y coefficients})\@.
Models with two doublets \( h_i \)
and \( p_i \) for each generation \( i = 1, 2, 3 \)
also are possible.

\section{Models with extended gauge groups
\label {Models with extended gauge groups}}

In most models of grand unification,
a Higgs mechanism breaks a simple 
gauge group \( G_u \) into the group 
of the standard model 
%\( SU_c(3) \otimes SU_w(2) \otimes U_Y(1) \)
at a unification energy \( E_u \)\@.
If this energy lies 
somewhat above \( 10^{15} \) GeV,
then proton decay,
proceeding through the exchange
of a single heavy gauge boson,  
is slow enough
not to have been seen in current 
experiments~\cite{Miura:2016krn}\@. 
A unification energy 
that high also lets the coupling
parameters of the subgoups \( SU_c(3) \), 
\( SU_w(2) \), and \( U_Y(1) \) of \( G_u \)  
run to values close 
to those observed  at TeV 
energies---at least if there's no new
relevant physics between \( 10^3 \) 
and \( 10^{15} \) GeV\@.
%\( 10^{34} \) years 
\par
Grand unification takes a different form
when the action of the model conserves parity
as in the models
of sections~\ref{A model for one generation}
\& \ref{A model for three generations}\@.
On the one hand, 
fermion fields and antifermion fields 
do not occur in the
same multiplets.  Thus
nucleon decay is intrinsically slow,
proceeding through the exchange
of three gauge bosons, and so
the unification energy \( E_u \)  
can be much lower than \( 10^{15} \) GeV, 
perhaps as low as a PeV\@.
On the other hand,
the energy of unification \( E_u \)
must be high enough that the 
three coupling parameters run long
enough to unify.
Two possibilities come to mind
depending upon whether there's
new physics between a TeV and \( E_u \)\@.
\par
Without new physics in that grand desert,
the energy of unification \( E_u \) and the mass \(M\)
of the heavy gauge bosons would be of the 
order of \( M \sim 10^{15} \) GeV or higher.
As we'll see presently, in simple extensions
of the model of this paper,
nucleon decay proceeds via the exchange of three
heavy gauge bosons.
The lifetime of the nucleon therefore rises
with the twelfth power of the ratio of the 
mass of the heavy gauge boson to that of the proton,
\( \tau_n \sim M^{12} / (\alpha^4_u m_p^{13}) \),
where \( \alpha_u \) is the fine-structure
constant of the unified theory.
The resulting nucleon lifetime
of more than \( 10^{150} \) years
would be too long for 
nucleon decay to be seen.
\par
If there is new physics below \(  10^{15} \) GeV,
then the simple group \( G_u \) might break twice.
For instance, it might break at an energy \( E_u \)
to \( SU(n) \otimes SU_w(2) \otimes \tilde G \)
and then break again to the group of the standard model
\( SU_c(3) \otimes SU_w(2) \otimes U_Y(1) \)
at a lower energy \( E_s \)\@.
The larger the integer \( n \), 
the faster the coupling parameter of \( SU(n) \)
runs between \( E_u \) and \( E_s\)\@.
\par
The existence 
of three generations of fermions
below a TeV may be a sign of
new physics below \(  10^{15} \) GeV\@.
One can imagine that at an energy \( E_u \)
the simple group breaks down to
\( SU(9) \otimes SU_w(2) \otimes G_1 \),
and then at \( E_s \)
this group breaks to
\( SU_c(3) \otimes SU_w(2) \otimes U_Y(1) \)\@.
Between \( E_u \) and \( E_s \),
the coupling parameters \( g_9 \) and 
\( g_2 \) run as~\cite{Weinberg1996.327}
\begin{equation}
   \begin{split}
   \mu \frac{d g(\mu)}{d\mu} 
   ={} & - \frac{g^3(\mu)}{4 \pi^2}
   \left( \frac{11}{12} C_1 - \frac{1}{3} C_2 \right)
   \end{split}
\end{equation}
in which \( C_1 = n \) for \( SU(n) \)
and \( C_2 = n_f/2 \) where
\( n_f \) is the number of fermions
in the representation of \( SU(n) \)\@.
For \( SU(9) \) there are two primary
multiplets \( U \) and \( D \) of 
three generations of three colors 
for a total of
nine quarks and similarly two secondary 
nonets \( U' \) and \( D' \)
of quarks, so \( n_{f, 9} = 4 \)\@.
For \( SU_w(2) \), there are nine quark
doublets, and three lepton doublets,
so \( n_{f, 2} = 12 \)\@.
Thus between \( E_u \) and \( E_s \),
the coupling parameters \( g_9 \) and 
\( g_2 \) run as
\begin{equation}
   \begin{split}
   \mu \frac{d g_9(\mu)}{d\mu} 
   ={} & - \frac{g^3_9(\mu)}{4 \pi^2}
   \left( \frac{11}{12} 9 - \frac{1}{3} \frac{4}{2} \right)
   = - \frac{g^3_9(\mu)}{4 \pi^2}
   \left( \frac{33}{4} - \frac{2}{3}  \right)
   = - \frac{91 \, g^3_9(\mu)}{48 \pi^2}
   \\
 \mu \frac{d g_2(\mu)}{d\mu} 
   ={} & - \frac{g^3_2(\mu)}{4 \pi^2}
   \left( \frac{11}{12} 2 - \frac{1}{3} \frac{12}{2} \right) 
   = - \frac{g^3_2(\mu)}{4 \pi^2}
   \left( \frac{11}{6}  -2 \right) 
   = \frac{g^3_2(\mu)}{24 \pi^2} 
   \end{split}
\end{equation}
which shows that \( SU_w(2) \)
is not asymptotically free
in this model.
Integrating, we get
\begin{equation}
   \begin{split}
   \frac{1}{g^2_9(E_s)} ={}&
   \frac{1}{g^2_9(E_u)}
   - \frac{91}{2 4 \pi^2}
   \log\left(\frac{E_u}{E_s}\right)
   \\
  \frac{1}{g^2_2(E_s)} ={}&
   \frac{1}{g^2_2(E_u)}
   + \frac{1}{12 \pi^2}
   \log\left(\frac{E_u}{E_s}\right) .
   \end{split}
\end{equation} 
The traces are 
\( \tr\left[(\thalf g_9 \lambda_3)^2\right] 
={} 6 g_9^2\)  and 
\( \tr\left[(\thalf g_2 \sigma_3)^2\right] 
={} 6 g_2^2\)\@.
So setting \( g^2_9(E_u) = g^2_2(E_u) \)
and subtracting, we find 
\begin{equation}
 \frac{1}{\alpha_2(E_s)} - \frac{1}{\alpha_9(E_s)} 
={} \frac{31}{2\pi} \log \left(\frac{E_u}{E_s}\right) .
\end{equation}
Thus
\begin{equation}
E_u ={} E_s  \, \exp\left[
\frac{2\pi}{31}\left(
\frac{1}{\alpha_2(E_s)} - \frac{1}{\alpha_9(E_s)}
\right) \right] .
\end{equation}
If \( E_s = 100 \) TeV where
\( 1/\alpha_2(E_s) - 1/\alpha_9(E_s) 
\sim 33  - 17 = 16 \)~\cite{Patrignani:2016xqp}, then
the higher energy scale is
\( E_u = 100 \, \exp(32\pi/31) = 2561 \) TeV
= 2.56 PeV\@.

\par
We can imagine putting the nonets \( U \) and \( D \)
of primary fermions and the triplets
\( E = (e, \mu, \tau) \) and 
\( N = ( \nu_e, nu_\mu, \nu_\tau) \) 
into the a multiplet \(F \) of dimension 24
\begin{equation}
F = \begin{pmatrix} 
U\\D\\E\\N
\end{pmatrix} .
\end{equation}
One would put the secondary fermions
into a similar 24-plet \(F'\)\@.
\par
I will assume that the group \( G_u \) 
has colored gauge bosons 
\( \vec J = ( \tcr{J}, \tcdg{J}, \tcb{J} ) \) 
that mediate \( \tcb{u} \to \nu +\tcb{J}^{2/3} \)
and other colored gauge bosons
\( \vec K = (\tcr{K}, \tcdg{K}, \tcb{K}) \)
that mediate 
\( \tcb{d} \to \nu + \tcb{K}^{-1/3} \)
and that a Higgs mechanism gives them masses
of at least a PeV\@.
I also assume that the cubic Yang-Mills coupling 
allows the process
\( \tcr{ J}^{2/3}+ \tcb{K}^{-1/3} \to
\tcdg{\bar K}^{1/3} \)\@.
In such grand unifications
of the model of section~\ref{A model for three generations},
the proton is unstable to decays like 
\( p \to \pi^+ + 3 \nu \) and \( p \to e^+ + 4 \nu \)
since the proton and the combinations \( \pi^+ + 3 \nu \)
and \( e^+ + 4 \nu \) all have
\( F = 3 \) and electric charge \({}+1\)\@.
These decays tend to be slow because
they involve three heavy gauge bosons as in
the process
\begin{equation}
   \begin{split}
   (\tcr{u}, \tcb{u}, \tcdg{d}) {}& \to
(\tcb{u}, \tcdg{d}) + \nu + \tcr{ J}^{2/3} 
\to \tcb{u} + \nu + \tcr{ J}^{2/3} 
+ \nu + \tcdg{K}^{-1/3} \\
{}& \to \tcb{u} + \nu + \nu + \tcb{\bar K}^{1/3}
\to \tcb{u} + \tcb{\bar d}+ \nu + \nu + \nu
\to \pi^+ + \nu + \nu + \nu .
\label {proton is unstable}
      \end{split}
\end{equation}
The neutron has \( F = 3 \) and 
charge zero.  Any state of three light neutrinos
also has \( F = 3 \) and charge zero.
So a neutron inside a nucleus can decay into three neutrinos,
\( n \to 3 \nu \)\@.
This decay also tends to be slow because
it also involves three heavy gauge bosons as in
the process
\begin{equation}
   \begin{split}
(\tcr{u}, \tcb{d}, \tcdg{d}) {}& \to
(\tcb{d}, \tcdg{d}) + \nu + \tcr{ J}^{2/3} 
\to  \tcdg{d} + \nu + \nu 
+ \tcr{ J}^{2/3}+ \tcb{K}^{-1/3} \\
{}& \to \tcdg{d} + \nu + \nu + \tcdg{\bar K}^{1/3}
\to \nu + \nu + \nu .
\label {neutron decay mode}
   \end{split}
\end{equation}
If the masses of these heavy gauge bosons
are \( M_J \) and \( M_K \),
then the lifetimes of the proton and 
of the nuclear neutron
in these models are proportional to 
\( m_J^8 m_K^4/(\alpha_u^6 m_p^{13}) \)
in which \( \alpha_u \) is the 
fine-structure constant of 
the unified theory.
\par
The lower bounds on nucleon partial lifetimes
are \(4.9 \times10^{26} \) years for 
\( n \to 3 \nu \)~\cite{Suzuki:1993zp},
\( 5.8 \times10^{29} \) years for 
\( n \to \) invisible~\cite{Araki:2005jt},
and \( 2.1 \times10^{29} \) years
for \( p \to \) invisible~\cite{PhysRevLett.92.102004}\@.
So the masses of the mediating gauge bosons
\( J \) and \( K \) 
should be a PeV or more.
This energy scale is 9 or 10
orders of magnitude lower
than the scale of traditional grand unification
because these models don't
put fermions and antifermions
in the same multiplets.
The running of masses and coupling constants
poses less of a fine-tuning problem
between 1 TeV and \( 10^4 \) TeV
than between 1 TeV and \(10^{12} \) TeV\@.
\par
Neutron decay might be seen in the SNO, KamLAND, 
Super-Kamiokande, and JUNO detectors and may
lurk in their recorded data.
The decay of a nucleon
in an \({}^{16}\)O nucleus would leave behind
an excited \( {}^{15}\)N\(^*\)
or \( {}^{15}\)O\(^*\) nucleus which
45\% of the time emits
a \(\gamma\) ray of 6--7 MeV~\cite{PhysRevLett.92.102004}\@.
In its ground state, an \({}^{15}\)O nucleus
has a half-life of 122\,s and
decays into a stable nucleus of \({}^{15}\)N,
a positron \( e^+ \), and
a neutrino \( \nu_e \)\@.
The energy of the positron can be as high as 1.732 MeV
with a mean energy of 735.28 keV~\cite{Epositron}\@.
\par
The present model requires the existence
of heavy mirror fermions with interactions
much like those of the known fermions.
They would form heavy positive
nuclei surrounded by shells
of heavy mirror electrons with
\( m_{e'} \gg m_e \)\@.
These mirror atoms would be
very small with Bohr radii
of the order of \( (\alpha_u m_{e'} )^{-1} \)\@.
The photon energy needed to excite
these atoms would be of the order
of \( m_{e'} \alpha^2 \),
and so a photon would need an energy 
in excess of 100 MeV to excite
one of these atoms.
These atoms are candidates for dark matter.
Because their masses would exceed 10 TeV,
their number density would be about 1000 times
lower than that of a 10 GeV \textsc{wimp}\@.
This low number density may be why physicists 
have not detected dark matter despite
its energy density being 5.4 times greater
than that of ordinary matter~\cite{Ade:2015xua}\@.
\par
The primary and secondary fermions both
interact through the gluons of \( SU_c(3) \),
so one might think that the nuclei 
of the heavy neutral atoms 
of the present model would interact
strongly with those of ordinary matter.
But at low energies nuclei scatter off other nuclei
by exchanging pions, and the
analog of a pion in 
heavy-nucleus--light-nucleus scattering
is a \textsc{wimp}y pion consisting of
a light quark and a heavy antiquark
or a light antiquark and a heavy quark.
A \textsc{wimp}y pion would have a mass in
excess of 1 TeV and so the cross-section
for heavy-nucleus--light-nucleus scattering
would be like that of a weak interaction.
The quarks of these putative dark-matter particles
interact with QCD, but their
interactions are weak
because the exchanged 
heavy pions are so massive.
Theories in which dark matter consists of
stealthy strongly interacting 
particles~\cite{Appelquist:2015zfa, *Appelquist:2015yfa} 
or strongly interacting massive particles 
(SIMPs)~\cite{Hochberg:2014kqa, *Hochberg:2015vrg}
have been developed\@.
Strongly interacting dark matter~\cite{Boddy:2014yra} 
broadens dark-matter cusps into 
cores~\cite{Rocha:2012jg, *Vogelsberger:2012ku}, 
as suggested by some observations, 
and so may explain the apparent paucity of heavy
dwarf galaxies around our galaxy~\cite{Zavala:2012us}\@.
\par
Some mechanism---perhaps initial conditions or
\(CP\)-violation---has created an excess of matter
over antimatter and possibly of dark matter
over dark antimatter.
The standard model does not have enough
\(CP\)-violation, but the CKM matrices
(\ref{mirror CKM matrices})
of the heavy fermions might.
If dark matter is composed of heavy quarks 
and leptons, then 
the symmetry
between the fermions and the
mirror fermions
may explain why the two excesses differ 
only by a factor of 5.4\@.

\section{Summary
\label {Summary}}

Although current research may change our understanding
of gravity, I have assumed in this paper that 
the action of a fundamental theory should be
invariant under general coordinate 
transformations so as to be compatible
with general relativity.  If this is so,
then the standard model should be extended 
to one whose action
is invariant under spatial reflection,
which is a simple coordinate transformation.  
This paper describes such a model. 
In the model,
the mean value in the vacuum of a pseudoscalar field
breaks parity.
This field and a scalar Higgs field
make the gauge bosons, 
the known fermions, and a set of mirror fermions 
suitably massive
while avoiding flavor-changing neutral currents
due to a novel kind of Yukawa alignment.
Because the action of the model
is invariant under spatial reflection,
the theory conserves quark-plus-lepton number
and has no anomalies
and no strong-\(CP\) problem.
The restoration of parity could occur at energies
as low as 10 TeV\@.
The model predicts heavy mirror fermions
which form heavy neutral mirror atoms which
are dark-matter candidates.
In some grandly unified extensions of 
the model, the scale of grand unification
can be as low as 2.5 PeV, which reduces
the fine-tuning problem, and nucleons 
slowly decay into pions,
antileptons,  and 
neutrinos in processes like \( p \to \pi^+ + 3 \nu \),
\( p \to e^+ + 4 \nu \), and
\( n \to 3 \nu \) that conserve fermion number
but violate \(B - L\)\@.

\begin{acknowledgments}
Conversations with Rouzbeh Allahverdi,
Andr{\'{e}} de
Gouv{\^{e}}a, Wick Haxton,
John Schwarz,
Shashank Shalgar, and Tim Tait
advanced this work as did
conversations with 
Robert Cooper, Franco Giuliani, 
Raphael Gobat, Michael Gold, Gary Herling, 
Kurt Hinterbichler, C.~W. Kim, 
John Matthews, and Anthony Zee and 
e-mail from Rouzbeh Allahverdi, 
Daniel Finley, Don Lichtenberg, and Howard Georgi.
I also wish to thank Jooyoung Lee for inviting
me to the Korea Institute for Advanced Study
where some of this paper was written.
\end{acknowledgments}
\bibliography{physics}

\end{document}